\documentclass[12pt]{iopart}
\usepackage{epsf}
\def\be{\begin{equation}}
\def\en{\end{equation}}                  
\begin{document}

\title[Simulating Colloid Dispersions]
{A Smooth Interface Method for Simulating Liquid Crystal Colloid Dispersions}

\author{Ryoichi Yamamoto\footnote[3]{To
whom correspondence should be addressed (ryoichi@scphys.kyoto-u.ac.jp)},
Yasuya Nakayama, and Kang Kim}

\address{Department of Physics, Kyoto University, Kyoto 606-8502, Japan
and PRESTO, Japan Science and Technology Agency, 
4-1-8 Honcho, Kawaguchi 332-0012, Japan.}

\begin{abstract}
A new method is presented for mesoscopic simulations of particle 
dispersions in liquid crystal solvents. 
It allows efficient first-principle simulations of the dispersions 
involving many particles with many-body interactions mediated by 
the solvents. 
Demonstrations have been performed for the aggregation of colloid 
dispersions in two-dimensional nematic and smectic-C* solvents
neglecting hydrodynamic effects, which will be taken into account
in the near future.

\end{abstract}

%Uncomment for PACS numbers title message
%\pacs{61.30.Cz, 61.30.Jf, 61.20.Ja}

% Uncomment for Submitted to journal title message
\submitto{\JPCM}

% Comment out if separate title page not required
%\maketitle

\section{Introduction}

Dispersions of small particles in host fluids such as 
colloid suspensions and emulsions are of considerable 
technological importance and appear often in our everyday life as
paints, foods, and drugs for example.
Many kinds of exotic interactions were found between particles 
mediated by the host fluids including 
screened Coulombic \cite{SC}, depletion \cite{SC}, 
fluctuation induced \cite{casimir}, 
and surface induced \cite{borstnik} forces.
A striking example occurs when spherical particles are
immersed in a liquid-crystal (LC) solvent in the nematic phase \cite{stark2}.
For a single particle, the orientation of the solvent molecules 
is distorted due to the anchoring of the solvent molecules at the
particle surface.
Extensive studies have been done on this effect, 
and several characteristic configurations of the nematic 
field around a spherical particle have been identified 
\cite{terentjev-sat,ramaswamy,terentjev-sim,stark,mondain,gu}.
When the strength of anchoring is increased so that 
normal anchoring is preferred, the solvent changes from 
quadrupolar to dipolar symmetries around the particle.
When multiple particles are immersed in the solvent, 
long-range anisotropic interactions are induced between particles 
due to elastic deformations of the nematic field
\cite{poulin,terentjev-the,lubensky,lev}.
The anisotropic interactions can have a pronounced effect not only 
on the local correlations of the particles \cite{poulin}, 
but also on their phase behavior 
\cite{raghunathan,terentjev-exp,loudet,jyamamoto,nazarenko,tyamamoto}
and on their mechanical properties \cite{terentjev-exp}.
It has been reported also that a similar, but somewhat different 
situation occurs when colloid particles are immersed in smectic-C* 
films \cite{cluzeau,patricio,pettey}.
The purpose of our project is to develop an efficient method
suitable for simulating colloid dispersions immersed in LC solvents
by using smooth interface between colloids and solvents.
Effects of hydrodynamics is omitted in the present paper
though it is very important \cite{care, fukuda2}.
That will be taken into account in the near future
also through the smooth interface \cite{NY01}.
The same method is applicable for charged colloid
dispersions where the charge density on colloid surface
is given by a smooth function \cite{KY01}.

\section{Mesoscopic Model}
 
Since analytical approaches for investigating complex materials 
such as LC colloid dispersions considered here are extremely
difficult, computer simulations 
are the most promising tool to investigate their static and 
dynamical properties.
In colloid dispersions, the host fluid molecules are much smaller and 
move much faster than the dispersed particles.
This enables us to use coarse grained mesoscopic variables, which 
should be described by the hydrodynamics, 
for the host fluids rather than treating them as fully microscopic 
objects \cite{MD,MD2}.
In the case of charged colloid suspensions for example, 
a mesoscopic method for the first principle simulations can be derived 
by treating the counter ions as a charge density \cite{lowen}. 
For the particle dispersions in LC solvents considered here, 
the mesoscopic free energy ${\cal F}$ of the system can be given by 
functionals of the director {\bf n}({\bf r}), a common direction 
on which solvent molecules are aligned on average with a constraint 
$|{\bf n}({\bf r})|=1$, for a given particle 
configuration $\{{\bf R}_1\cdots{\bf R}_N\}$:
\begin{equation}
{\cal F}({\bf n}({\bf r});\{{\bf R}_1\cdots{\bf R}_N\})
={\cal F}_{el} + {\cal F}_s
=\int d{\bf r}f_{el}({\bf r})+\oint dSf_{s}(S),
\label{total}
\end{equation}
where
\begin{equation}
f_{el}({\bf r})
=\frac{1}{2}\left[
 K_1(\nabla\cdot{\bf n})^2
+K_2({\bf n}\cdot\nabla\times{\bf n})^2
+K_3\{{\bf n}\times(\nabla\times{\bf n})\}^2
\right]
\label{elastic}
\end{equation}
is the Frank free energy \cite{frank} which presents
the elastic energy density of the nematic solvent at ${\bf r}$ and 
the surface free energy
\begin{equation}
f_{s}(S)=\frac{W}{2}[1-({\bf n}\cdot\mbox{\boldmath$\nu$})^2]
\label{surface}
\end{equation}
controls anchoring of the LC solvent at the 
surface element $S$ of the particle.
The coefficients $K_1$, $K_2$, and $K_3$ are the splay, twist, and 
bend elastic constants, respectively, and
the single constant approximation ($K=K_1=K_2=K_3$) reduces 
Eq.(\ref{elastic}) to the simpler form
\begin{equation}
f_{el}({\bf r})
=\frac{K}{2}\left[(\nabla\cdot{\bf n})^2
+(\nabla\times{\bf n})^2
\right].
\label{elastic1}
\end{equation}
$W$ is the surface anchoring constant, 
and \mbox{\boldmath$\nu$} is the unit
vector normal to the colloid surface \cite{terentjev-sim,stark}.
The saddle-splay elastic term \cite{stark} is not considered here.
The integral in the first term on the right hand side of
Eq.(\ref{total}) runs over the whole 
solvent volume excluding the particles, 
and that in the second term runs over all solvent-particle interfaces.
A simple scaling argument tells us ${\cal F}_{el}\propto Ka^{d-2}$
and ${\cal F}_{s}\propto Wa^{d-1}$ with $a$ and $d$ being 
the particle radius and the system dimension, respectively, thus
the state of the dispersion should be controlled by the ratio 
${\cal F}_{s}/{\cal F}_{el}\propto Wa/K$.
Although this type of free energy functional is sufficient for 
single particle problems, 
it is not useful for simulating colloid dispersions involving 
many particles using molecular dynamics (MD) or Brownian type methods
because the coupling between solvent 
and the particles is given implicitly by limiting the integration space 
in both ${\cal F}_{el}$ and ${\cal F}_{s}$.
This produces mathematical singularities at the interface when one 
calculates the force, 
${\bf f}_n^{PS}=-\partial {\cal F}/\partial {\bf R}_n$, acting on each 
particle mediated by the LC solvents.
Calculation of the force is crucial for performing efficient
simulations of many particle systems.
Another serious problem of this type of functional 
is that in order to give correct boundary conditions at the 
particle-solvent interface, one has to use appropriate coordinates 
for performing grid-based numerical simulations rather than the 
usual Cartesian coordinates.
This is generally difficult for particles with non-spherical shapes or 
for systems involving many particles even when each particle has 
a spherical shape.
Also this makes the use of the periodic boundary condition difficult, 
which is a fatal situation for simulating bulk materials.

\begin{figure}[t]
\centerline{
\epsfxsize=2.6in\epsfbox{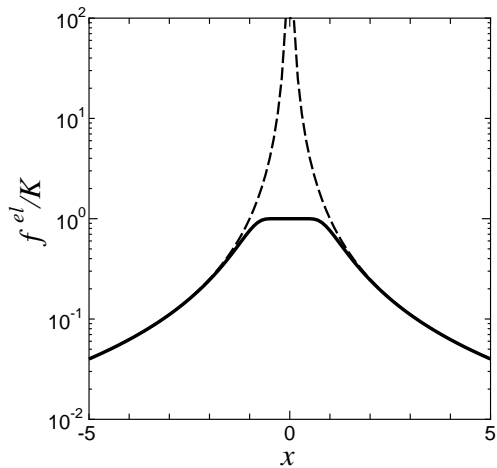}}
\caption{%\protect\narrowtext 
The elastic free energy density $f_{el}(x)$ around a point defect
at $x=0$.
The dashed line is from Eq.(\ref{elastic1}), $f_{el}\simeq K/x^2$,
and the solid line is from Eq.(\ref{ryn2}) or (\ref{rys2})
with $R_c=1$,
$f_{el}\simeq K\tanh(1/x^2)$.
}
\label{fig1}
\end{figure}

To overcome these problems, we have modified 
Eqs.(\ref{total}), (\ref{surface}), and (\ref{elastic1}) by
using a smooth interface between the solvent and the particles 
so that the coupling is given explicitly in the integrand 
through the interface \cite{RY}.
The new free energy functional we propose has the form
\begin{eqnarray}
{\cal F}({\bf n}({\bf r});\{{\bf R}_1&\cdots&{\bf R}_N\})\nonumber\\
&=&\int d{\bf r}\left[1-\sum_{i=1}^N\phi_i({\bf r})\right]
f^{el}({\bf r})
+\int d{\bf r}\sum_{i=1}^N\xi(\nabla\phi_i)^2f^{s}_i({\bf r})
\label{ryn1}
\end{eqnarray}
with 
\begin{eqnarray}
f^{el}({\bf r})&=&\frac{K}{4R_c^2}
\tanh\left[R_c^2(\nabla_\alpha (n_\beta n_\gamma))^2\right]
\label{ryn2}\\
f^{s}_i({\bf r})
&=&\frac{W}{2}
\left[1-\left(\frac{\nabla\phi_i}{|\nabla\phi_i|}\cdot {\bf n}\right)^2\right]
\label{ryn3}
\end{eqnarray}
for nematic solvent and 
\begin{eqnarray}
f^{el}({\bf r})&=&\frac{K}{2R_c^2}
\tanh\left[R_c^2((\nabla\cdot{\bf n})^2+ (\nabla\times{\bf n})^2)\right]
\label{rys2}\\
f^{s}_i({\bf r})
&=&{W}\left[1-\frac{\nabla\phi_i}{|\nabla\phi_i|}\cdot {\bf n}\right]
\label{rys3}
\end{eqnarray}
for smectic-C* solvent.
The summation convention is used for $\alpha,\beta,\gamma\in x,y,z$.
The explicit form of the interfacial profile $\phi_n$ between dispersed
particles and solvents and its spatial derivatives $\nabla_\alpha\phi_n$ 
are given by
\be
\phi_n({\bf r})
=\frac{1}{2}\left(\tanh\frac{a-|{\bf r}-{\bf
R}_n|}{\xi}+1\right)
\label{profile1}
\en
and
\be
\nabla_\alpha\phi_n({\bf r})
=-\frac{r_\alpha-R_{n,\alpha}}{2\xi|{\bf r}-{\bf
R}_n|}\cosh^{-2}\frac{a-|{\bf r}-{\bf R}_n|}{\xi},
\label{profile2}
\en
respectively with the particle radius $a$ and the interface thickness $\xi$.
Note that a set of Eqs.(\ref{ryn1})-(\ref{profile2}) reduces to 
Eqs.(\ref{total}), (\ref{surface}), and (\ref{elastic1}) 
for the limit $R_c,\xi\rightarrow0$.
A similar idea for using smooth interface was used
to treat the hydrodynamic forces acting on particles dispersed 
in simple liquids \cite{tanaka}, recently.
For the nematic case, $f_{el}$ is given by a function of a symmetric 
second-rank tensor $q_{\alpha\beta}=n_\alpha n_\beta$
rather than the director ${\bf n}$ itself to take into account the 
symmetry of the nematic director $+{\bf n}\leftrightarrow-{\bf n}$
automatically.
The semi-empirical functional form 
$\frac{1}{R_c^2}\tanh[R_c^2\cdots]$
is applied in Eqs.(\ref{ryn2}) and (\ref{rys2}) to avoid 
the mathematical 
divergence of the elastic free energy density at the defect centers
and to limit its value to $\Delta f_{el}\sim K/R_c^2$ and  
replace its core size with $R_c$ as shown in Figure \ref{fig1}.
Another way to avoid the divergence would be to use  
the Landau--de Gennes type free energy with an order parameter 
$
Q_{\alpha\beta}({\bf r})=Q({\bf r})q_{\alpha\beta}({\bf r}),
$
but this requires a prohibitively small lattice spacing near the 
defect points \cite{fukuda}.

%%%%% Fig.1 %%%%%%%%%%%%%%%%%%

\section{Numerical method}

\subsection{First-principle Simulation}

The simulation procedure is as follows. 
\begin{enumerate}
\item For a given particle
configuration $\{{\bf R}_1\cdots{\bf R}_N\}$, obtain
the interface profile $\phi_n({\bf r})$ by Eq.(\ref{profile1}). 
Then we can calculate the stable (or meta-stable) nematic configuration
${\bf n}^{(0)}({\bf r})$ which should 
satisfy the equilibrium condition
\be
\left.\frac{\delta{\cal F}}{\delta
{\bf n}({\bf r})}\right|_{{\bf n}({\bf r})={\bf n}^{(0)}({\bf r})}=0
\label{fderiv}
\en
under the director constraint $|{\bf n}({\bf r})|=1$.
One can perform this by numerical iterations such as the steepest descent or 
the conjugate gradient method.
Although the time evolutions of ${\bf n}({\bf r})$ should be determined
by the hydrodynamic laws, we here assume
${\bf n}({\bf r})$ follows adiabatically after $\{{\bf R}_1\cdots{\bf R}_N\}$ 
for the sake of simplicity.

\item Once ${\bf n}^{(0)}({\bf r})$ is obtained, 
the force acting on each particle mediated by the nematic solvents 
follows directly from the Hellmann-Feynman theorem,
\begin{eqnarray}
{\bf f}_n^{PS}(\{{\bf R}_1&\cdots&{\bf R}_N\})
\equiv-\frac{\partial {\cal F}({\bf n}^{0}({\bf r});\{{\bf R}_1\cdots{\bf R}_N\})}{\partial {\bf R}_n}\\
&=&\frac{K}{4R_c^2}
\int d{\bf r}
\frac{\partial\phi_n}{\partial {\bf R}_n}
\tanh\left[R_c^2(\nabla_\alpha (n_\beta^{0}n_\gamma^{0}))^2\right]\nonumber\\
&&+W\xi\int d{\bf r}\frac{\partial(\nabla_\alpha\phi_n)}{\partial {\bf R}_n}(\nabla_\beta\phi_n)n_\alpha^{0} n_\beta^{0}
\label{fne}\\
%\end{eqnarray}
%\begin{eqnarray}
&=&\frac{K}{2R_c^2}
\int \hspace{-1mm}d{\bf r}
\frac{\partial\phi_n}{\partial {\bf R}_n}
\tanh\left[R_c^2((\nabla\hspace{-0.5mm}\cdot\hspace{-0.5mm}{\bf n}^0)^2+ (\nabla\hspace{-1mm}\times\hspace{-0.5mm}{\bf n}^0)^2)\right]\nonumber\\
%\tanh\left[R_c^2(\nabla_\alpha (n_\beta^{0}n_\gamma^{0}))^2\right]
&&+W\xi\int \hspace{-1mm}d{\bf r}\frac{\partial(\nabla_\alpha\phi_n)}{\partial {\bf R}_n}
\left[\frac{\nabla_\alpha\phi_n}{|\nabla\phi_n|}\nabla_\beta\phi_n n^0_\beta+|\nabla\phi_n|n^0_\alpha\right],
\label{fsm}
\end{eqnarray}
where Eqs.(\ref{fne}) and (\ref{fsm}) are for the nematic and smectic-C*
      solvents, respectively.
These forms are very convenient because one can compute both
$\partial\phi_n/\partial {\bf R}_n$ and 
$\partial(\nabla_\alpha\phi_n)/\partial {\bf R}_n$ at any time
since $\phi_n$ is an analytical function of ${\bf R}_n$.

\item Finally, update the particle positions according to
appropriate equations of motion such as 
\be
m_n\frac{d^2{\bf R}_n}{dt^2}
={\bf f}_n^{PP}+{\bf f}_n^{PS}+{\bf f}_n^H+{\bf f}_n^R ,
\label{motion}
\en
where ${\bf f}_n^{PP}$ is the force due to direct particle-particle 
interactions (hard or soft sphere for instance), ${\bf f}_n^{H}$
and ${\bf f}_n^{R}$ are the hydrodynamic and random forces.
Repeating the steps (i)$\sim$(iii) enables us to perform 
first-principles mesoscopic simulations for the dispersions containing 
many particles without neglecting many-body interactions.

\end{enumerate}

\section{Results}

\subsection{Director configurations around a single particle}

We have performed simple demonstrations for two-dimensional (2D) 
LC colloid dispersions to test the performance of our procedure.
The demo system has $100\times100$ lattice sites in a square box with 
a linear length $L=100$.
Other physical parameters are chosen rather arbitrarily as 
$R_c=1$, $a=5$, and $\xi=2$, where the unit of length is the lattice
spacing $l$.
Since the director configurations in 2D can be expressed by a single scalar
field $\theta({\bf r})$, the tilt angle of the director against the
horizontal ($x$-) direction,
Eq.(\ref{fderiv}) then reduces to
\be
%\left.
\frac{\delta{\cal F}}{\delta\theta({\bf r})}
%\right|_{\theta({\bf r})=\theta^{(0)}({\bf r})}
=\frac{\partial n_{\alpha}({\bf r})}
{\partial\theta({\bf r})}\frac{\delta{\cal F}}
{\delta n_{\alpha}({\bf r})}
=0 .
\en
%with $q_{xx}=\cos^2\theta-1/2$, $q_{yy}=\sin^2\theta-1/2$,
%and $q_{xy}=q_{yx}=\cos\theta\sin\theta$.
The boundary condition is fixed at $\theta({\bf r})=0$ at the 
edge of the box to avoid rotations of the reference frame.

%%%%% Fig.2 %%%%%%%%%%%%%%%%%%
\begin{figure}[t]
\centerline{
\epsfxsize=2.5in\epsfbox{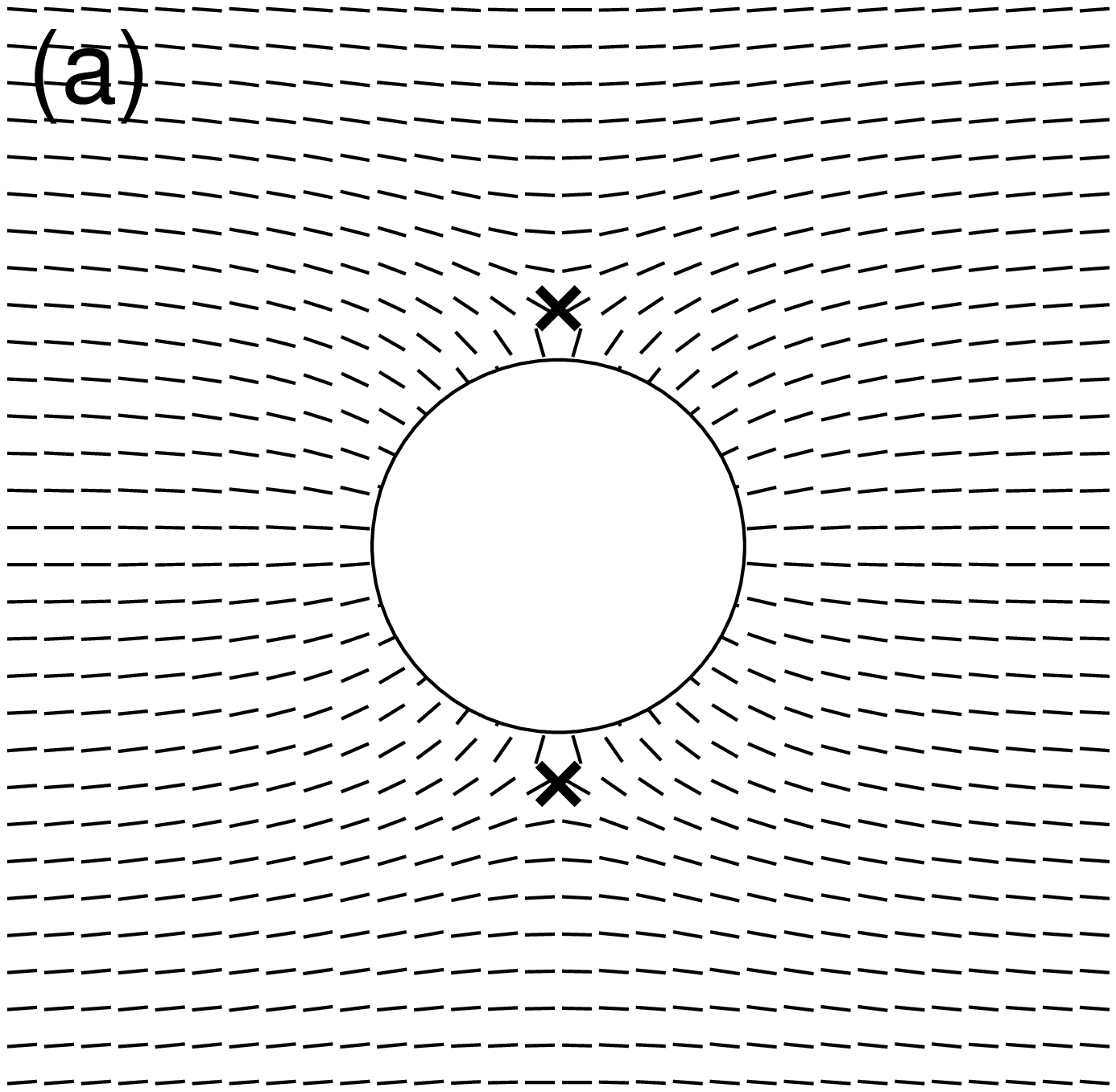}\hspace{10mm}
\epsfxsize=2.5in\epsfbox{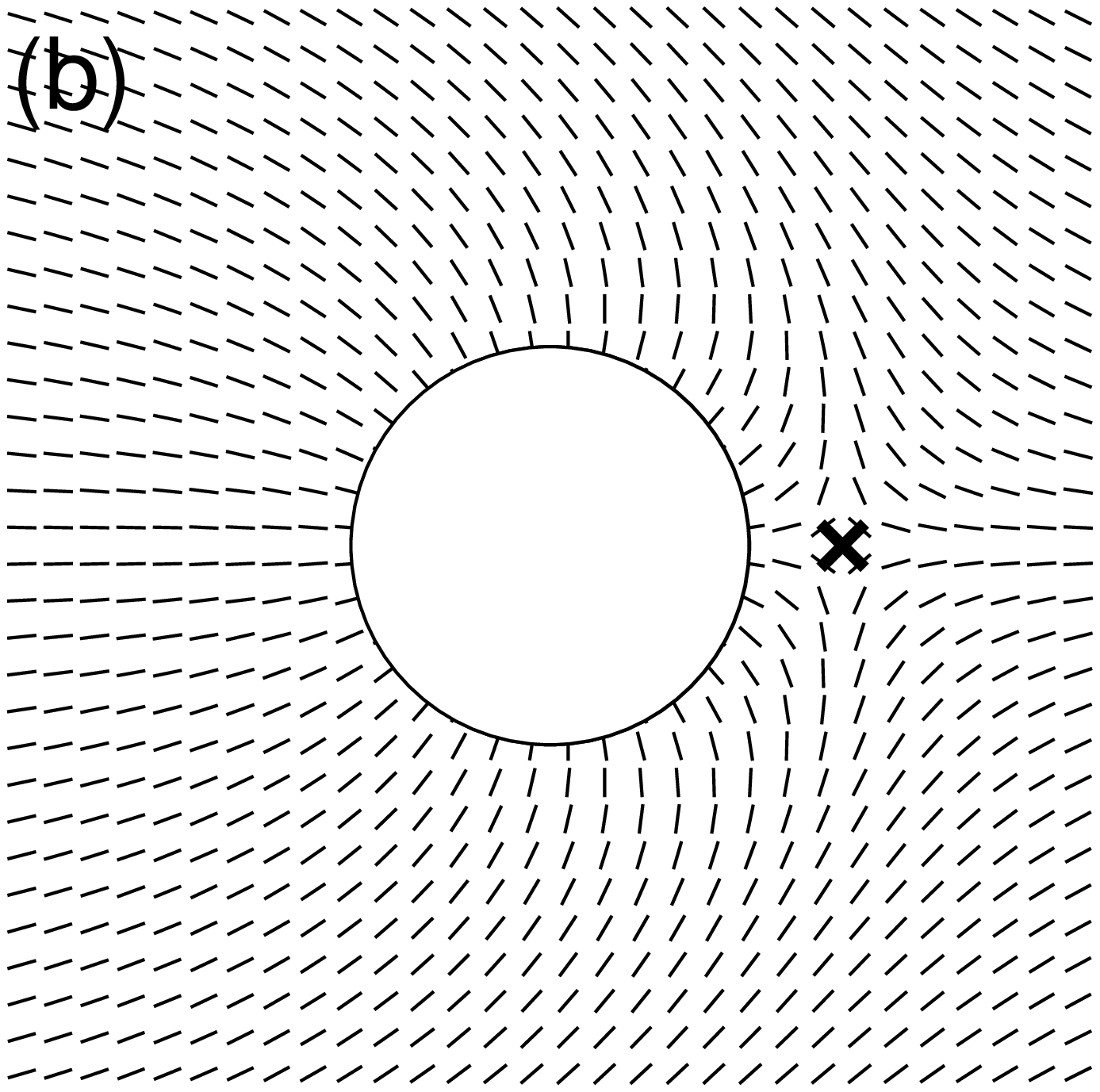}}\vspace{3mm}
\caption{\protect%\narrowtext 
(a) Director configuration around a single particle immersed in 
nematic solvent with strong anchoring condition $Wa/K=4$.
The crosses show two $-1/2$ charge point defects and 
the obtained configuration has a quadrupolar character.
(b) Director configuration around a single particle immersed in
smectic-C* solvent with $Wa/K=5$.
The cross shows $-1$ charge point defect and the obtained 
configuration has a dipolar character.
The white disks indicate the colloid particles. 
Only $9$\% of the total system is shown for display purpose.
}
\label{fig2}
\end{figure}

We first calculated the stable director configurations around a single 
particle immersed in 2D nematic and smectic-C* solvents and showed them 
in Figure \ref{fig2} (a) and (b), respectively.
For the nematic solvent (a), the particle is accompanied by two $-1/2$ 
charge point defects with a strong anchoring condition $Wa/K=4$.
The distance between the defects and the particle center is about $1.2a$,
which is similar to the analytic value $1.236a$ \cite{fukuda}, 
and the director configuration around the single particle 
possess quadrupolar symmetries.
This would correspond to the Saturn ring configuration observed 
in three dimensional (3D) systems.
Although in principle particles can be accompanied by one $-1$ 
charge hedgehog defect in 2D as well as in 3D, 
such configurations are unstable in the present 
2D system since the elastic penalty 
of having $m$ point defects with charge $c$ scales as $mKc^2$.
This was directly confirmed by recent simulations with perfect 
normal anchoring \cite{fukuda} and also by our simulations.
For the smectic-C* solvent (b) on the other hand,
the particle is accompanied by one $-1$ charge point defect with 
an anchoring condition $Wa/K=5$.
The distance between the defect and the particle center is about 
$1.4a$ which is again similar to the experimental value 
$1.4a\pm0.1a$ \cite{cluzeau} and the analytic one $\sqrt{2}a$ \cite{pettey}.
Note that the director configuration around the particle 
possess dipolar symmetries similar to the experiment \cite{cluzeau}.

\subsection{aggregation of colloids in LC solvents}

\begin{figure}[tbhp]
\centerline{\epsfxsize=5.in\epsfbox{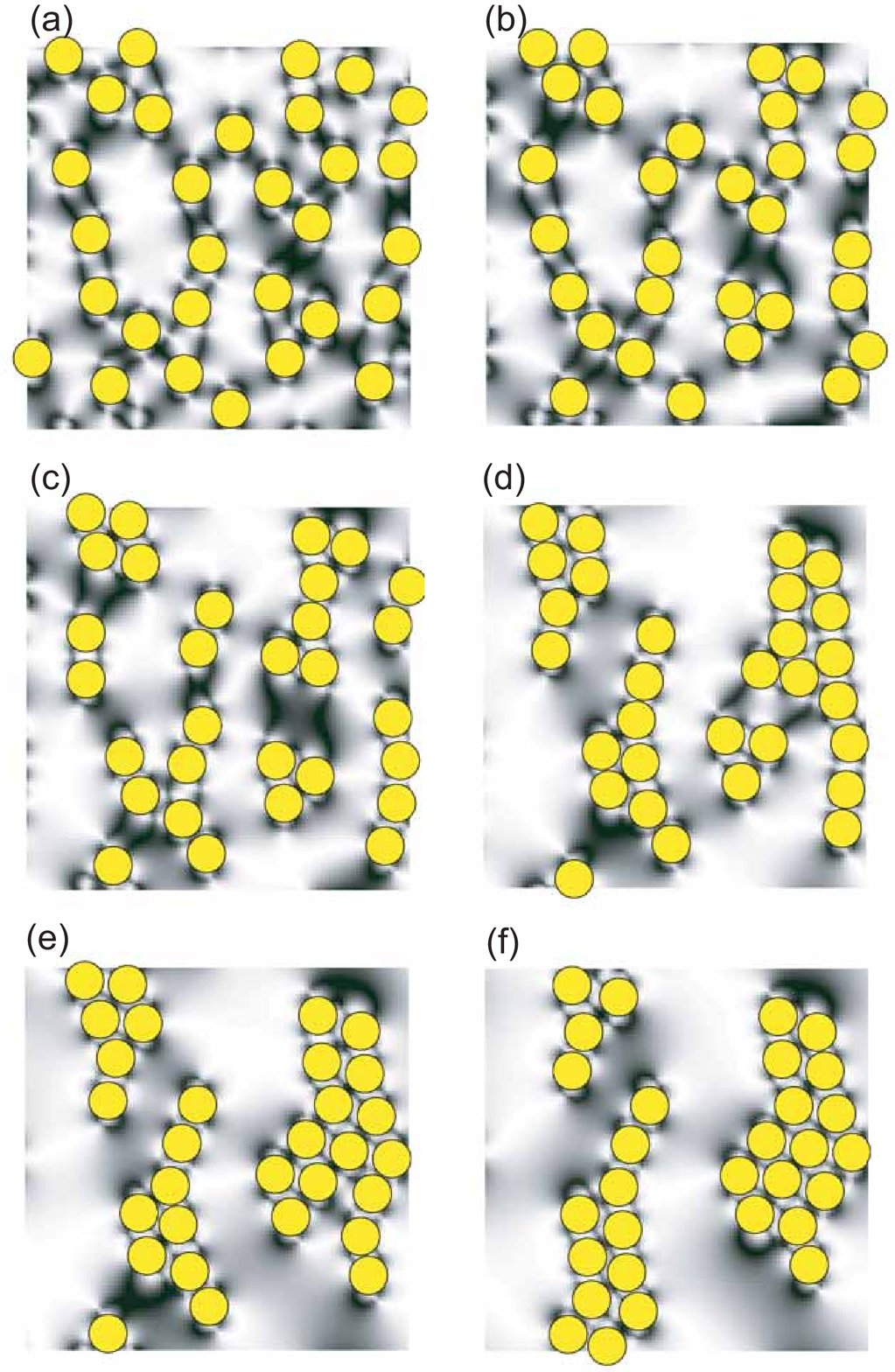}}\vspace{0mm}
\centerline{\epsfxsize=2.6in\epsfbox{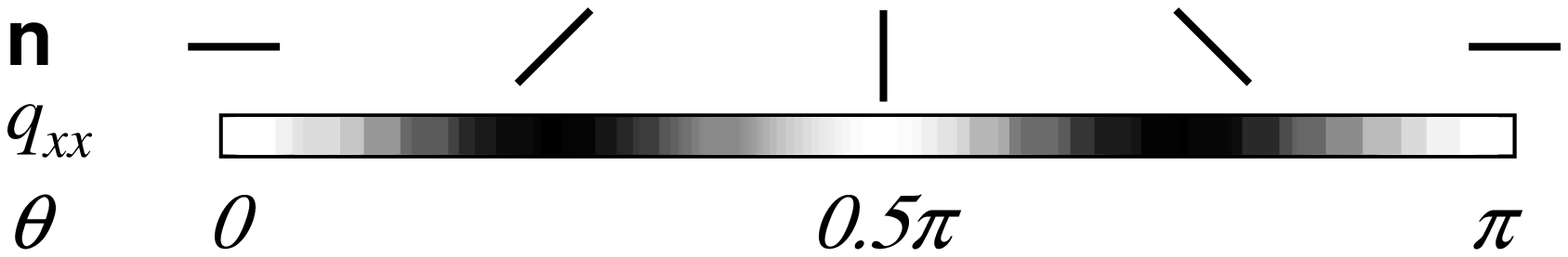}}
\caption{\protect%\narrowtext 
The aggregation and ordering process of colloid particles 
after the solvent exhibit the isotropic ($K=0$) to nematic ($K=1$) 
phase transition; 
(a) $t=0$, (b) $t=0.4$, (c) $t=1.6$, (d) $t=6.4$, (e) $t=16$, and (f) $t=40$.
Each particle is accompanied by two $-1/2$ charge point defects.
Darkness presents the value of $q_{xx}^2$.
Black and white correspond to $q_{xx}^2=0$ and $0.25$, respectively.
Those correspond also to $\theta=0.25\pi,0.75\pi$ and $\theta=0,0.5\pi,\pi$ 
as shown in the gradation map.
}
\label{fig3}
\end{figure}

\begin{figure}[tbhp]
\centerline{\epsfxsize=5.in\epsfbox{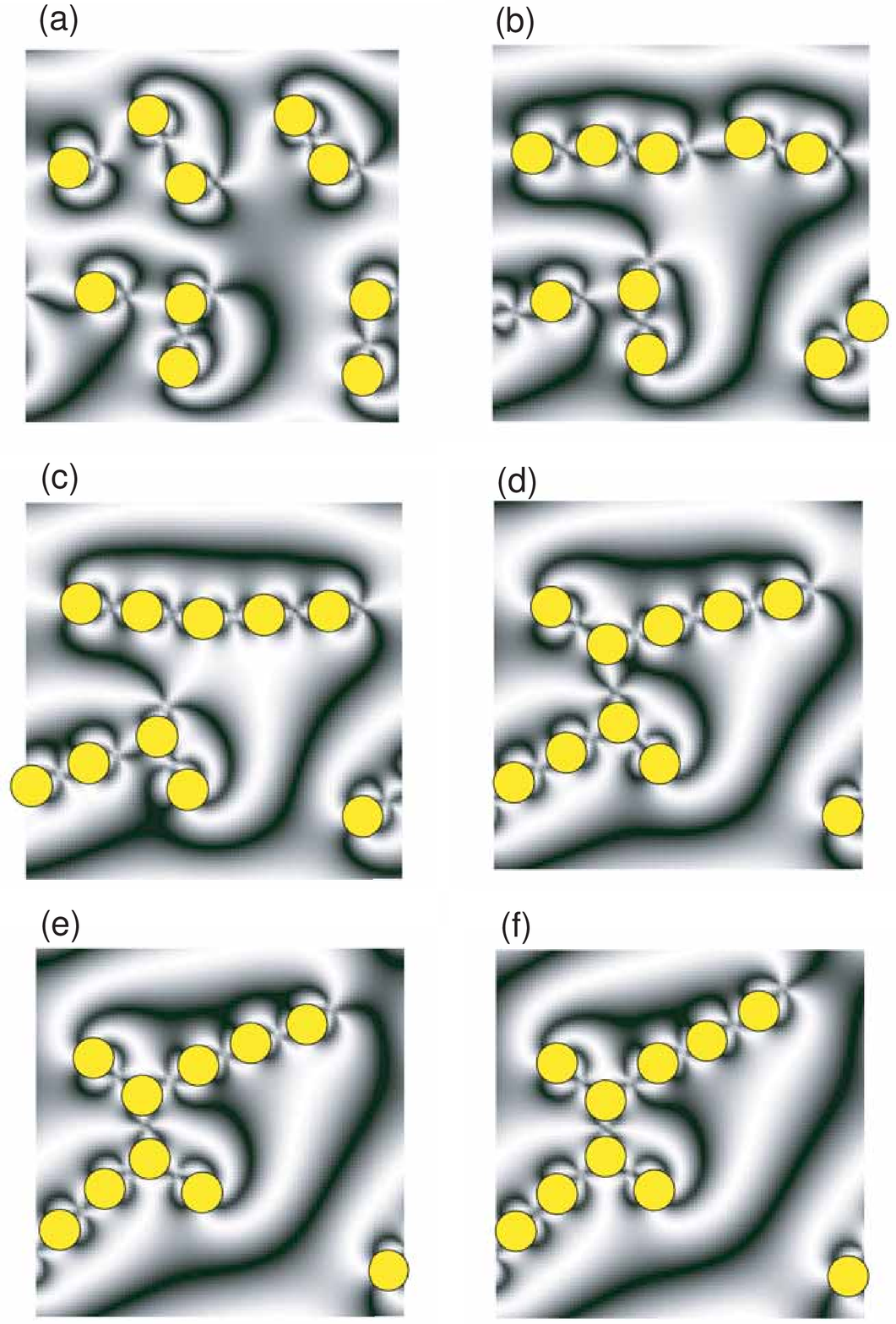}}\vspace{0mm}
\centerline{\epsfxsize=2.6in\epsfbox{map.eps}}
\caption{\protect%\narrowtext 
The aggregation and ordering process of colloid particles 
after the solvent exhibit the isotropic ($K=0$) to smectic-C* ($K=1$) 
phase transition; 
(a) $t=0$, (b) $t=0.25$, (c) $t=1$, (d) $t=4$, (e) $t=10$, (f) $t=25$. 
Each particle is accompanied by one $-1$ charge point defect.
Darkness presents the value of $q_{xx}^2$.
Black and white correspond to $q_{xx}^2=0$ and $0.25$, respectively.
Those correspond also to $\theta=0.25\pi,0.75\pi$ and $\theta=0,0.5\pi,\pi$ 
as shown in the gradation map.
}
\label{fig4}
\end{figure}

We have simulated the aggregation and ordering process of $30$ and $10$
colloid particles in nematic and smectic-C* solvents, respectively, 
after the isotropic to nematic or isotropic to smectic-C* transition 
occurred.
Here we used the periodic boundary condition and set $Wa/K=4$ (nematic) 
or $5$ (smectic C*).
Other parameters are the same as in the previous single particle case.
The simulation was performed starting from a random particle
configuration which is a typical configuration when the solvent is in
the isotropic phase ($K=0$). 
We then set $K=1$ and calculated ${\bf f}_n^{PS}$ according to 
the present procedure. 
The particle configurations were updated by 
numerically solving the steepest descent equation,
\be
\zeta\frac{d {\bf R}_n}{dt}={\bf f}_n^{PS}+{\bf f}_n^{PP},
\en
which is obtained by simply substituting 
$d^2{\bf R}_n/dt^2=0$, ${\bf f}_n^R=0$, and 
${\bf f}_n^H=-\zeta d {\bf R}_n/dt$ in Eq.(\ref{motion}).
$\zeta=1$ is a friction constant and thus the off-diagonal components 
of the hydrodynamic interaction were not considered.
Here we obtain 
\be
{\bf f}_n^{PP}=-\frac{\partial E_{PP}}{\partial{\bf R}_n}
\en
from the repulsive part of the Lennard-Jones potential, 
\be
E_{PP}=0.4\sum_{n=1}^{N-1}\sum_{m=n+1}^N\left[\left(
\frac{2a}{|{\bf r}_n-{\bf r}_m|}\right)^{12}
-\left(\frac{2a}{|{\bf r}_n-{\bf r}_m|}\right)^{6}+\frac{1}{4}\right]
\en
truncated at the minimum distance $|{\bf r}_n-{\bf r}_m|=2^{7/6}a$,
to avoid the particles overlapping each other within 
the colloid radius $\simeq a$.
Snapshots from the present simulation are shown in Figures \ref{fig3} 
and \ref{fig4} for dispersions in nematic and smectic C* solvemts, 
respectively.
In Figure \ref{fig3}, the particles are forming ordered clusters 
due to the quadrupolar attractive interaction among them.
In Figure \ref{fig4} on the other hand, the particles are forming 
string-like clusters due to the dipolar attractive interaction 
among them.
Similar results have been obtained recently by experiments \cite{cluzeau}.
Note that only up to two particle simulations have been done 
so far \cite{stark} and simulations of more than two particles 
would be extremely difficult or almost impossible even for 2D
systems by means of other methods ever proposed.

%%%%%  Fig.3 %%%%%%%%%%%%%%%%%%

%%%%%  Fig.4 %%%%%%%%%%%%%%%%%%

\section{Concluding Remarks}

In summary, we have developed a powerful simulation method 
to investigate colloid dispersions interacting via solvents.
We proposed a free energy functional which is suitable for MD type 
simulations. 
The following modifications have been made to the original Frank 
free energy functional.
\begin{enumerate}
\item The coupling between the nematic solvent and particles at the
interfaces is introduced explicitly through a smooth interface so 
that we can analytically calculate the force acting
on each particle mediated by the host by taking derivatives 
of the free energy according to the particle positions.
\item The value of the free energy density is bounded semi-empirically
to avoid a mathematical divergence in the defect centers.
\end{enumerate}
We have performed demonstrations for 2D dispersions 
and confirmed that the method works quite well for systems which
contain defects.
Applications of this method to 3D systems should have no theoretical 
difficulties, but require somewhat heavier computation.
This should allow the simulation of the chaining of the particles
caused by the possible dipolar symmetry of the nematic configurations 
around a single particle.
Although we have shown only simple demonstrations of the method by
performing simulations of the 2D system in this letter, simulations 
with physically more interesting situations such as systems with 
non-circular particles, asymmetric particle pairs with different 
particle size, or particles with non-normal anchoring 
as well as more realistic simulations in 3D systems are possible.
Efforts for taking into account the hydrodynamic effects are now 
underway by using smooth interface instead of imposing hydrodynamic 
boundary conditions at the colloid interface.

\section*{Acknowledgements}

RY thanks Professor J.P. Hansen, Professor H. L\"owen, and 
Dr. E. Terentjev for helpful discussions.
A part of the present calculations have been carried out at 
the Human Genome Center, 
Institute of Medical Science, University of Tokyo.

\appendix
\section{numerical implementation for 2D nematic}
Here, the superscripts $i,j$ ($= 1,2,\cdots,L$)
denote positions on a 2D lattice ($L\times L$).

\begin{equation}
{\cal F}_{el}=\frac{K}{4R_c^2}\sum_{i,j}(1-\sum_{n=1}^N\phi_{n}^{i,j})
\tanh\epsilon^{i,j}
\end{equation}

\begin{eqnarray}
\epsilon^{i,j}=\frac{R_c^2}{2l^2}\left[
 (q_{\alpha\beta}^{i,j}-q_{\alpha\beta}^{i-1,j})^2\right.
&+&(q_{\alpha\beta}^{i+1,j}-q_{\alpha\beta}^{i,j})^2\nonumber\\
&+&\left.(q_{\alpha\beta}^{i,j}-q_{\alpha\beta}^{i,j-1})^2
+(q_{\alpha\beta}^{i,j+1}-q_{\alpha\beta}^{i,j})^2
\right]
\end{eqnarray}
\begin{eqnarray}
q_{\alpha\beta}^{i,j}&=&n_\alpha^{i,j} n_\beta^{i,j}
-\delta_{\alpha\beta}/2
\end{eqnarray}
\begin{equation}
{\cal F}_{s}=\frac{W\xi}{2}\sum_{i,j}\sum_{n=1}^N\left[
\frac{1}{2}(\nabla_\alpha\phi_{n}^{i,j})^2-(\nabla_\alpha\phi_{n}^{i,j})(\nabla_\beta\phi_{n}^{i,j})q_{\alpha\beta}^{i,j}\right]
\end{equation}

\begin{eqnarray}
\frac{\partial {\cal F}}{\partial \theta^{i,j}}
&&=\frac{\partial {\cal F}}{\partial q_{\alpha\beta}^{i,j}}\frac{\partial q_{\alpha\beta}^{i,j}}{\partial \theta^{i,j}}
=\frac{\partial {\cal F}}{\partial q_{xx}^{i,j}}\frac{\partial q_{xx}^{i,j}}{\partial \theta^{i,j}}
+2\frac{\partial {\cal F}}{\partial q_{xy}^{i,j}}\frac{\partial q_{xy}^{i,j}}{\partial \theta^{i,j}}
+\frac{\partial {\cal F}}{\partial q_{yy}^{i,j}}\frac{\partial q_{yy}^{i,j}}{\partial \theta^{i,j}}\nonumber\\
&&=2\sin\theta^{i,j}\cos\theta^{i,j}\left(
\frac{\partial {\cal F}}{\partial q_{yy}^{i,j}}
-\frac{\partial {\cal F}}{\partial q_{xx}^{i,j}}
\right)
+2(\cos^2\theta^{i,j}-\sin^2\theta^{i,j})
\frac{\partial {\cal F}}{\partial q_{xy}^{i,j}}
\end{eqnarray}

\begin{equation}
\frac{\partial {\cal F}}{\partial q_{\alpha\beta}^{i,j}}
=\frac{\partial {\cal F}_{el}}{\partial q_{\alpha\beta}^{i,j}}
+\frac{\partial {\cal F}_{s}}{\partial q_{\alpha\beta}^{i,j}}
\end{equation}

\begin{eqnarray}
\frac{\partial {\cal F}_{el}}{\partial q_{\alpha\beta}^{i,j}}
=&\frac{K}{4l^2}
&\left[(1-\sum_{n=1}^{N}\phi_{n}^{i,j})\cosh^{-2}\epsilon^{i,j}(4q_{\alpha\beta}^{i,j}-q_{\alpha\beta}^{i-1,j}-q_{\alpha\beta}^{i+1,j}-q_{\alpha\beta}^{i,j-1}-q_{\alpha\beta}^{i,j+1})\right.\nonumber\\
&+&(1-\sum_{n=1}^{N}\phi_{n}^{i-1,j})\cosh^{-2}\epsilon^{i-1,j}(q_{\alpha\beta}^{i,j}-q_{\alpha\beta}^{i-1,j})\nonumber\\
&+&(1-\sum_{n=1}^{N}\phi_{n}^{i+1,j})\cosh^{-2}\epsilon^{i+1,j}(q_{\alpha\beta}^{i,j}-q_{\alpha\beta}^{i+1,j})\nonumber\\
&+&(1-\sum_{n=1}^{N}\phi_{n}^{i,j-1})\cosh^{-2}\epsilon^{i,j-1}(q_{\alpha\beta}^{i,j}-q_{\alpha\beta}^{i,j-1})\nonumber\\
&+&\left.(1-\sum_{n=1}^{N}\phi_{n}^{i,j+1})\cosh^{-2}\epsilon^{i,j+1}(q_{\alpha\beta}^{i,j}-q_{\alpha\beta}^{i,j+1})\right]
\end{eqnarray}

\begin{equation}
\frac{\partial {\cal F}_{s}}{\partial q_{\alpha\beta}^{i,j}}
=-\frac{W\xi}{2}\sum_{n=1}^{N}\nabla_\alpha\phi_{n}^{i,j}\nabla_\beta\phi_{n}^{i,j}
\end{equation}

\begin{eqnarray}
{\bf f}_n^{PS}
&=&\frac{K}{4R_c^2}\sum_{i,j}
\frac{\partial\phi_n^{i,j}}{\partial {\bf R}_n}\tanh\epsilon^{(0)i,j}
+W\xi\sum_{i,j}\frac{\partial(\nabla_\alpha\phi_n^{i,j})}{\partial {\bf R}_n}(\nabla_\beta\phi_n^{i,j})q_{\alpha\beta}^{(0)i,j} .
\end{eqnarray}

\section{numerical implementation for 2D smectic-C*}

\begin{equation}
{\cal F}_{el}=\frac{K}{2R_c^2}\sum_{i,j}(1-\sum_{n=1}^N\phi_{n}^{i,j})
\tanh\epsilon^{i,j}
\end{equation}

\begin{eqnarray}
\epsilon^{i,j}=
\frac{R_c^2}{2l^2}\left[
(n_{\alpha}^{i,j}-n_{\alpha}^{i-1,j})^2\right.
&+&(n_{\alpha}^{i+1,j}-n_{\alpha}^{i,j})^2\nonumber\\
&+&(n_{\alpha}^{i,j}-n_{\alpha}^{i,j-1})^2
+\left.(n_{\alpha}^{i,j+1}-n_{\alpha}^{i,j})^2
%\right.\nonumber\\
%\left.+(n_{x}^{i+1,j}-n_{x}^{i-1,j})(n_{y}^{i,j+1}-n_{y}^{i,j-1})
\right]
\end{eqnarray}

\begin{equation}
{\cal F}_{s}=W\xi\sum_{i,j}\sum_{n=1}^N\left[
(\nabla\phi_{n}^{i,j})^2-|\nabla\phi_{n}^{i,j}|\nabla\phi_{n}^{i,j}\cdot {\bf n}^{i,j}
\right]
\end{equation}

\begin{eqnarray}
\frac{\partial {\cal F}}{\partial \theta^{i,j}}
&&=\frac{\partial {\cal F}}{\partial n_{\alpha}^{i,j}}\frac{\partial n_{\alpha}^{i,j}}{\partial \theta^{i,j}}
=\frac{\partial {\cal F}}{\partial n_{x}^{i,j}}\frac{\partial n_{x}^{i,j}}{\partial \theta^{i,j}}
+\frac{\partial {\cal F}}{\partial n_{y}^{i,j}}\frac{\partial n_{y}^{i,j}}{\partial \theta^{i,j}}\nonumber\\
&&=-n_y^{i,j}\left(\frac{\partial {\cal F}_{el}}{\partial n_{x}^{i,j}}+\frac{\partial {\cal F}_s}{\partial n_{x}^{i,j}}\right)
+n_x^{i,j}\left(\frac{\partial {\cal F}_{el}}{\partial n_{y}^{i,j}}+\frac{\partial {\cal F}_s}{\partial n_{y}^{i,j}}\right)
\end{eqnarray}

\begin{eqnarray}
\frac{\partial {\cal F}_{el}}{\partial n_{\alpha}^{i,j}}
=&\frac{K}{2l^2}&
\left[(1-\sum_{n=1}^{N}\phi_{n}^{i,j})\right.\cosh^{-2}\epsilon^{i,j}(4n_{\alpha}^{i,j}-n_{\alpha}^{i-1,j}-n_{\alpha}^{i+1,j}-n_{\alpha}^{i,j-1}-n_{\alpha}^{i,j+1})\nonumber\\
&+&(1-\sum_{n=1}^{N}\phi_{n}^{i-1,j})\cosh^{-2}\epsilon^{i-1,j}(n_{\alpha}^{i,j}-n_{\alpha}^{i-1,j})\nonumber\\
&+&(1-\sum_{n=1}^{N}\phi_{n}^{i+1,j})\cosh^{-2}\epsilon^{i+1,j}(n_{\alpha}^{i,j}-n_{\alpha}^{i+1,j})\nonumber\\
&+&(1-\sum_{n=1}^{N}\phi_{n}^{i,j-1})\cosh^{-2}\epsilon^{i,j-1}(n_{\alpha}^{i,j}-n_{\alpha}^{i,j-1})\nonumber\\
&+&\left.(1-\sum_{n=1}^{N}\phi_{n}^{i,j+1})\cosh^{-2}\epsilon^{i,j+1}(n_{\alpha}^{i,j}-n_{\alpha}^{i,j+1})\right]
\end{eqnarray}

\begin{equation}
\frac{\partial {\cal F}_{s}}{\partial n_{\alpha}^{i,j}}
=-W\xi\sum_{n=1}^{N}|\nabla\phi_{n}^{i,j}|\nabla_\alpha\phi_{n}^{i,j}
%[(\nabla_\alpha\phi_{n}^{i,j})n_{\alpha})]
\end{equation}

\begin{eqnarray}
{\bf f}_n^{PS}
&=&\frac{K}{2R_c^2}\sum_{i,j}
\frac{\partial\phi_n^{i,j}}{\partial {\bf R}_n}\tanh\epsilon^{(0)i,j}\nonumber\\
&+&W\xi\sum_{i,j}
%\left[(\nabla_\alpha\phi_n^{i,j})n_\alpha^{i,j}\right]
\frac{\partial(\nabla_\alpha\phi_n^{i,j})}{\partial {\bf R}_n}
\left(|\nabla\phi_{n}^{i,j}|n_\alpha^{i,j}
+\frac{\nabla\phi_{n}^{i,j}\cdot {\bf n}^{i,j}}{|\nabla\phi_{n}^{i,j}|}\nabla_\alpha\phi_{n}^{i,j}\right)
\end{eqnarray}

\section*{References}
%\begin{references}

%\end{references}
%\end{document}

\end{document}